\documentclass[final,5p,times,twocolumn,nopreprintline]{elsarticle}
\biboptions{sort&compress}
\journal{Physics Letters B}
\usepackage[utf8]{inputenc}
\usepackage[T1]{fontenc}

\usepackage{amsmath}
\usepackage{amssymb}
\usepackage{xcolor}
\usepackage{graphicx}
\usepackage{microtype}

\usepackage{braket}
\usepackage{balance}

\usepackage[colorlinks=true,citecolor=blue,linkcolor=blue,urlcolor=blue]{hyperref}
\usepackage{orcidlink}

\newcommand{\imsrg}{\mbox{IMSRG}}
\newcommand{\imsrgtwo}{\mbox{IMSRG(2)}}

\newcommand{\MeV}{\text{MeV}}

\newcommand{\fm}{\text{fm}}

\newcommand{\pb}{\ensuremath{{}^{208}\mathrm{Pb}}}
\newcommand{\ca}{\ensuremath{{}^{48}\mathrm{Ca}}}

\newcommand{\nn}{\ensuremath{\text{NN}}}
\newcommand{\nnn}{\ensuremath{\text{3N}}}

\newcommand{\vnn}{\ensuremath{V_\nn}}
\newcommand{\vnnn}{\ensuremath{V_\nnn}}

\newcommand{\dnnlogo}{\ensuremath{\Delta\mathrm{NNLO}_\mathrm{GO}}}
\newcommand{\arthuisem}{\ensuremath{\mathrm{1.8/2.0}~\mathrm{(EM7.5)}}}

\newcommand{\rchtwo}{\ensuremath{R_\mathrm{ch}^2}}
\newcommand{\rchfour}{\ensuremath{R_\mathrm{ch}^4}}

\newcommand{\rptwo}{\ensuremath{R_\mathrm{p}^2}}
\newcommand{\rpfour}{\ensuremath{R_\mathrm{p}^4}}

\newcommand{\rntwo}{\ensuremath{R_\mathrm{n}^2}}

\newcommand{\rskin}{\ensuremath{R_\mathrm{skin}}}
\newcommand{\apv}{\ensuremath{A_\mathrm{PV}}}
\newcommand{\fch}{\ensuremath{F_\mathrm{ch}}}
\newcommand{\fchint}{\ensuremath{F^{\rm intr}_{\rm ch}}}
\renewcommand{\fchint}{\ensuremath{F_{\rm ch}^{\rm int}}}

\newcommand{\hw}{\ensuremath{\hbar\omega}}

\newcommand{\ethreemax}{\ensuremath{E_\mathrm{3max}}}

\begin{document}

\title{Ab initio computations of the fourth-order charge density moments of $^{48}$Ca and $^{208}$Pb}

\author[tsuk]{T.~Miyagi\orcidlink{0000-0002-6529-4164}}
\ead{miyagi@nucl.ph.tsukuba.ac.jp}
\author[olcf,ornl]{M.~Heinz\orcidlink{0000-0002-6363-0056}}
\ead{heinzmc@ornl.gov}
\author[tud,emmi,mpik]{A.~Schwenk\orcidlink{0000-0001-8027-4076}}
\ead{schwenk@physik.tu-darmstadt.de}

\address[tsuk]{Center for Computational Sciences, University of Tsukuba, 1-1-1 Tennodai, Tsukuba 305-8577, Japan}
\address[olcf]{National Center for Computational Sciences, Oak Ridge National Laboratory, Oak Ridge, TN 37831, USA}
\address[ornl]{Physics Division, Oak Ridge National Laboratory, Oak Ridge, TN 37831, USA}
\address[tud]{Technische Universit\"at Darmstadt, Department of Physics, 64289 Darmstadt, Germany}
\address[emmi]{ExtreMe Matter Institute EMMI, GSI Helmholtzzentrum f\"ur Schwerionenforschung GmbH, 64291 Darmstadt, Germany}
\address[mpik]{Max-Planck-Institut f\"ur Kernphysik, Saupfercheckweg 1, 69117 Heidelberg, Germany}

\begin{abstract}

Neutron skins of neutron-rich nuclei connect nuclei with the matter in neutron stars.
High-precision measurements of nuclear charge densities to extract higher-order moments are proposed to be sensitive to neutron radii and skin thicknesses.
We investigate the charge density of $^{48}$Ca and $^{208}$Pb, leading candidates for such studies, with ab initio nuclear structure calculations.
We find strong correlations between the fourth-order charge density moment $R_\mathrm{ch}^4$ and the charge and neutron radii,
allowing us to predict $R_\mathrm{ch}^4$ for $^{48}$Ca and $^{208}$Pb.
We find a substantially weaker correlation between the fourth-order charge density moment and the neutron skin, limiting the ability of high-precision electron scattering to determine the neutron skin in a model-independent manner.

\end{abstract}

\begin{keyword}
neutron skin \sep electron scattering \sep ab initio nuclear theory

\end{keyword}

\maketitle

\section{Introduction}

Neutron skins of neutron-rich nuclei provide crucial insights into the properties of neutron-rich nucleonic matter, constraining the equation of state of neutron star matter around saturation density~\cite{Steiner2005PR, Thiel2019JPG}.
This drives efforts to infer the neutron skin thicknesses of systems like \ca{} and \pb{}, using parity-violating electron scattering and weak scattering~\cite{Adhikari2021,Adhikari2022,AtzoriCorona2023WeakScatteringRn}, coherent pion photoproduction~\cite{Tarbert2013PRL_Pb208PiProduction}, proton scattering~\cite{Zenihiro2010, Tamii2011PRL, Birkhan2017PRL}, anti-proton capture~\cite{Trzcinska2001PRL,Klos2007,PUMA2022ngr}, and high-precision electron scattering~\cite{Kurasawa2021PTEP_R4Measurement},
and the theoretical work supporting the analyses and interpretations of these experiments~\cite{Brown2000PRL,Horowitz2001PRL,Reinhard2010PRC,Fattoyev2010PRC,Piekarewicz2012PRC,Hebeler:2014ema,Roca-Maza2015PRC,Hagen2016NP_Ca48Skin,Payne2019PRC,Miller2019PRC,Kurasawa2019,Reinhard2020PRC,Hu2022NP_Pb208,Kurasawa2022PTEP, Fearick:2023lyz,Arthuis2024arxiv_LowResForces,Heinz2024inprep_MuToEResponses,Tsaran2025arxiv}.

\begin{figure*}[t!]
    \centering
    \includegraphics{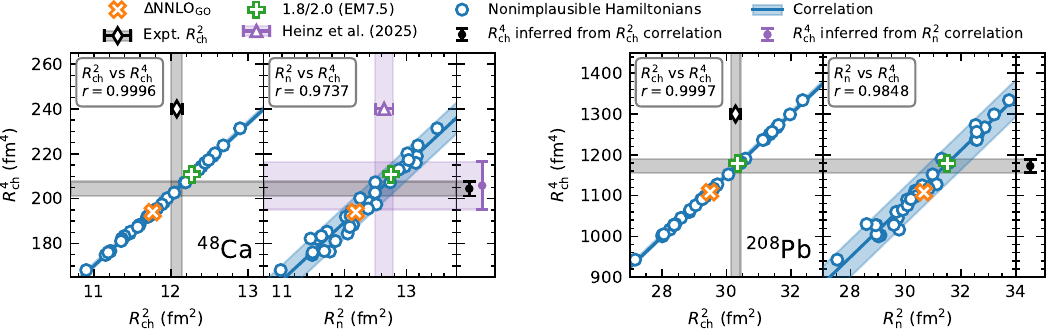}
\caption{Predictions of \rchtwo{}, \rchfour{}, and \rntwo{} (for \ca{} and \pb{} on the left and right, respectively) using the \dnnlogo{}~\cite{Jiang2020PRC_DN2LOGO} and \arthuisem{}~\cite{Arthuis2024arxiv_LowResForces} Hamiltonians and the 34 nonimplausible Hamiltonians from Ref.~\cite{Hu2022NP_Pb208}. We show best fit lines to the correlations between \rchfour{}, \rchtwo{}, and \rntwo{} for the nonimplausible Hamiltonians, each with a conservative error band. The Pearson correlation coefficient $r$ for each correlation is given in the top left. Using these correlations and experimental determinations of \rchtwo{} for \ca{}~\cite{Noel2024JHEP_ChargeDensities} and \pb{}~\cite{Kurasawa2021PTEP_R4Measurement} and a theoretical prediction of \rntwo{}~\cite{Heinz2024inprep_MuToEResponses}, we provide predictions for \rchfour{} of \ca{} and \pb{}.}
\label{fig:rch4_correlation}
\end{figure*}

While parity-violating electron scattering gives direct insights into the weak densities of the nuclei studied, resolving the small parity-violating asymmetry \apv{} makes these experiments challenging.
The CREX~\cite{Adhikari2022} and PREX~\cite{Adhikari2021} experiments on \ca{} and \pb{} have successfully measured the skin thicknesses of both systems, but with different implications for the neutron matter equation of state~\cite{Reed2021PRL,Reinhard2021PRL,Yuksel2022PLB}.
This tension is not fully understood, motivating both remeasurement of \apv{} by MREX~\cite{Schlimme2024MESA} and further theoretical analyses to infer the skin thickness~\cite{Reinhard2022PRL,Mondal2022PRC,Miyatsu2023PLB,Roca-Maza2025PRL,Reed:2025ccn}.

On the other hand, high-precision electron scattering to measure the charge form factor \fch{} is sensitive to the neutron density, in addition to other contributions.
Specifically, experimental efforts at ULQ2 at Tohoku University aim to determine \fch{} at very low momentum transfer $q$ to extract the fourth-order moment of the charge distribution \rchfour{}, which may be analyzed to infer the neutron radius and the neutron skin thickness~\cite{Kurasawa2019,Kurasawa2021PTEP_R4Measurement,Kurasawa2022PTEP}.
Such inferences have relied on phenomenological mean-field calculations, yielding model-dependent results for neutron radii~\cite{Kurasawa2021PTEP_R4Measurement}.

Now ab initio calculations using nuclear Hamiltonians rooted in quantum chromodynamics can predict the structure of nuclei as heavy as \pb{}~\cite{Hu2022NP_Pb208,Hebeler2023PRC_JacobiNO2B,Miyagi2024PRL_MagMoment,Arthuis2024arxiv_LowResForces, Door2024arxiv_YbBoson}.
In this Letter, we study the charge form factors of \ca{} and \pb{} using ab initio calculations to establish microscopic correlations between moments of the proton and neutron distributions and the moments of the charge distribution \rchtwo{} and \rchfour{} that can be extracted from high-precision electron scattering experiments.

\section{Ab initio computations}

Ab initio nuclear structure calculations~\cite{Hergert2020FP_AbInitioReview} start from nuclear Hamiltonians
\begin{equation}
H = T_\mathrm{int} + \vnn + \vnnn\,,
\end{equation}
with the intrinsic kinetic energy $T_\mathrm{int}$
and two- and three-nucleon potentials \vnn{} and \vnnn{}, respectively.
Such Hamiltonians are rooted in quantum chromodynamics through effective field theories (EFTs) of the strong interaction~\cite{Epelbaum2009RMP_ChiralEFTReview, Machleidt2011PR_ChiralEFTReview}
and optimized to nucleon-nucleon scattering data, properties of few-nucleon systems, and occasionally also selected medium-mass nuclei~\cite{Entem2003PRC_EM500, Hebeler2011PRC_SRG3NFits, Ekstrom2015PRC_N2LOsat, Epelbaum2015PRL_EKMN4LO,Jiang2020PRC_DN2LOGO, Arthuis2024arxiv_LowResForces}.
Nuclear Hamiltonians are intrinsically uncertain, due to both truncations of the EFT expansion and fitting of unknown low-energy couplings (LECs) to experimental data.

In this work, we employ two selected chiral EFT Hamiltonians optimized for reproduction of ground-state energies and charge radii of medium-mass and heavy nuclei, the \dnnlogo{} (with cutoff 394\:MeV) and \arthuisem{} Hamiltonians~\cite{Jiang2020PRC_DN2LOGO, Arthuis2024arxiv_LowResForces}.
We additionally employ an ensemble of 34 so-called ``nonimplausible Hamiltonians'' constructed at next-to-next-to-leading order in chiral EFT with Delta isobar degrees of freedom~\cite{Hu2022NP_Pb208}.
This ensemble is very conservative in its construction, incorporating maximal LEC variations within EFT uncertainties.
It is reflective of the worst case uncertainty of nuclear Hamiltonians and allows us to systematically explore this uncertainty. 

\begin{figure*}[t!]
    \centering
    \includegraphics{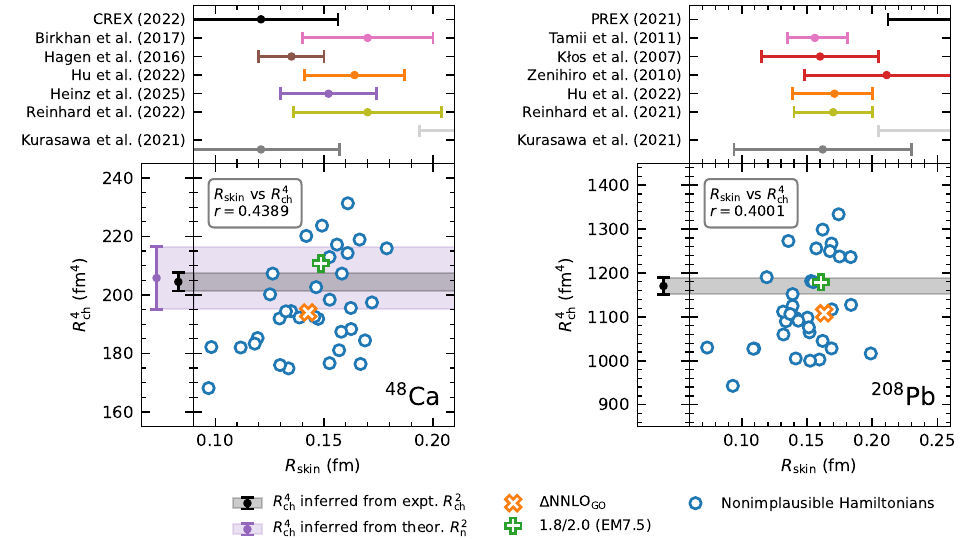}
\caption{Predictions of \rskin{} and \rchfour{} (for \ca{} and \pb{} on the left and right, respectively) using the \dnnlogo{}~\cite{Jiang2020PRC_DN2LOGO} and \arthuisem{}~\cite{Arthuis2024arxiv_LowResForces} Hamiltonians and the 34 nonimplausible Hamiltonians from Ref.~\cite{Hu2022NP_Pb208}. We compare with our determinations of \rchfour{} from Fig.~\ref{fig:rch4_correlation}, which are based on experimental \rchtwo{} measurements~\cite{Kurasawa2021PTEP_R4Measurement,Noel2024JHEP_ChargeDensities} and a theoretical \rntwo{} prediction for \ca{}~\cite{Heinz2024inprep_MuToEResponses}.
We also compare with literature values for \rskin{} (above), from parity-violating electron scattering (CREX/PREX, black)~\cite{Adhikari2021, Adhikari2022}, electric dipole responses (Birkhan et al./Tamii et al., pink)~\cite{Tamii2011PRL, Birkhan2017PRL}, hadronic probes (K{\l}os et al, Zenihiro et al., red)~\cite{Klos2007,Zenihiro2010}, ab initio predictions (Hagen et al., brown; Hu et al., orange; Heinz et al., purple)~\cite{Hagen2016NP_Ca48Skin, Hu2022NP_Pb208, Heinz2024inprep_MuToEResponses}, nonrelativistic density-functional theory (DFT) predictions using the SV-min functional (Reinhard et al., olive)~\cite{Reinhard2021PRL, Reinhard2022PRL}, and inferences from experimental data based on relativistic and nonrelativistic DFT correlations (Kurasawa et al., light and dark gray, respectively)~\cite{ Kurasawa2021PTEP_R4Measurement}.}
\label{fig:skin_correlation}
\end{figure*}

For each Hamiltonian, we solve the many-body Schrödinger equation using in-medium similarity renormalization group (\imsrg{})~\cite{Hergert2016PR_IMSRG} starting from a Hartree-Fock reference state.
Many-body methods like the IMSRG are approximate, but systematically improvable, with mild polynomial scaling in system size that enables calculations of medium-mass and heavy nuclei~\cite{Hagen2014RPP_CCReview, Hergert2016PR_IMSRG, Tichai2020FP_MBPTReview, Soma2020FP_SCGCReview}. 
The \imsrgtwo{} approximation, where the IMSRG is truncated at the normal-ordered two-body level, has been demonstrated to be precise for ground-state properties of nuclei~\cite{Heinz2025}.

We compute the ground states of \ca{} and \pb{} using the \imsrgtwo{} for all the Hamiltonians mentioned above.
For our calculations of \pb{}, we expand all matrix elements of our Hamiltonian in a spherical harmonic oscillator basis of 15 major shells with a frequency of $\hw = 12\:\MeV$.
We additionally truncate our \nnn{} potentials in the three-body basis $\ket{pqr}$, keeping only states with $e_p + e_q + e_r \leq \ethreemax = 28$ ($e = 2n + l$).
These truncations have been demonstrated to be sufficient for model-space convergence for the ground state of \pb{}~\cite{Hu2022NP_Pb208,Hebeler2023PRC_JacobiNO2B}.
For our calculations of \ca{} we employ instead a basis of 13 major shells with a larger frequency $\hbar\omega=16\:\MeV$ that is more optimal for convergence in medium-mass nuclei~\cite{Hoppe2021PRC_NAT}.

We compute charge form factors from the ground-state expectation values of the charge form factor operator:
\begin{equation}
\begin{aligned}
F_{\rm ch}(&q^{2}) = e \sum_{i=1}^{A} 
\left\{
G^{E}_{i}(q^{2}) \left[1 - \frac{q^{2}}{8m^{2}} \right] j_{0}(qr_{i}) 
\right. \\ &  \left.
- \frac{q^{2}}{2m^{2}} \left[G^{M}_{i}(q^{2}) - \frac{1}{2} G^{E}_{i}(q^{2}) \right] (\vec{\ell}_{i} \cdot \vec{\sigma}_{i}) \frac{j_{1}(qr_{i})}{qr_{i}}
\right\}.
\end{aligned}
\end{equation}
Here $q$ is the magnitude of the momentum transfer, $e$ is the elementary electric charge, $G^{E}_{i}(q^{2})$ and $G^{M}_{i}(q^{2})$ are the $i$th nucleon's electric and magnetic form factors, $\vec{\ell}_{i}$ is the orbital angular momentum operator, $\vec{\sigma}_{i}$ is the spin Pauli matrix, and $j_{l}(x)$ is the order $l$ spherical Bessel function of the first kind.
In this work we adopt $G^{E}_{i}(q^{2})$ and $G^{M}_{i}(q^{2})$ from Ref.~\cite{Ye2018}.
We do not consider the contributions of two-body currents to the charge form factor.
These have been shown to be of the same order of magnitude as spin-orbit corrections~\cite{Krebs:2020pii}, which contribute $< 1\%$ and $< 0.1\%$ to \rchtwo{} in \ca{} and \pb{}, respectively.

We make the form factor translationally invariant by leveraging the Gaussian factorization of the center-of-mass wave function, originally established for coupled-cluster calculations~\cite{Hagen2009PRL_CoMCorrections} and also confirmed in IMSRG calculations~\cite{Hergert2016PR_IMSRG,Heinz2024inprep_MuToEResponses}: 
\begin{equation}
\fchint(q^{2}) = e^{q^{2}b^{2}_{\rm cm}/4} F_{\rm ch}(q^{2})\,,
\end{equation}
where $b_{\rm cm}$ is the oscillator length of the center-of-mass motion estimated from the center-of-mass kinetic energy expectation value.

\section{Charge density moments}

The moments of nuclear densities may be computed in several different ways.
One can construct and evaluate operators for the individual contributions, as is typically done for the charge radius squared or the neutron radius squared~\cite{Hagen2016NP_Ca48Skin,Heinz2025}.
Once one considers the fourth moment \rchfour{}, such operator expressions become complicated and unwieldy~\cite{Kurasawa2019, Miyagi2025}.
It is instead easier to extract such moments from the charge form factor directly.
This can be done by computing $\fchint(q^{2})$ for a large set of momentum transfers $q$, performing a Fourier transform to get the charge density $\rho_\mathrm{ch}(r)$, and then computing the desired moments directly via integration:
\begin{equation}
\label{eq:Rn-dens}
\langle R_\mathrm{ch}^n \rangle = \frac{\int_0^\infty dr\, r^{n+2} \rho_\mathrm{ch}(r)}{\int^{\infty}_{0} dr\, r^{2}\rho_{\rm ch}(r)}\,.
\end{equation}
Alternatively, one can compute them via derivatives of $\fchint(q^{2})$ as $q^2\to0$, which requires relatively few evaluations of $\fchint(q^{2})$ at small $q$.
For instance, \rchfour{} may be computed as
\begin{equation}
\rchfour{} = \left. \frac{60}{\fchint(0)} \frac{d^2 \fchint(q^{2})}{(dq^2)^2} \right|_{q^{2}=0}\,.
\end{equation}

We compute \rchtwo{} and \rchfour{} by evaluating $\fchint(q^2)$ for 11 equally spaced points from $q^{2}= 0$ to $1$ fm$^{-2}$, constructing a Gaussian process interpolator, and taking derivatives as $q^2\to0$.
Gaussian process interpolation provides robust second derivatives of \fchint{} and allows the uncertainties related to the interpolation and differentiation to be quantified (see \ref{app:moments} for details).
We find that for all of our calculations, the estimated uncertainty on \rchtwo{} and \rchfour{} due to the Gaussian process interpolation is less than 0.001~$\fm^2$ and 0.3~$\fm^4$ for \ca{} (0.005~$\fm^2$ and 1.8~$\fm^4$ for \pb{}), respectively.

\section{Fourth-order moments of \ca{} and \pb{}}

It is well known that predictions of various moments of nuclear charge, weak, point-proton, and point-neutron densities are correlated.
These correlations arise due to the common many-body physics and nuclear Hamiltonians at play, limiting how much, for example, proton and neutron densities can vary independently.
Figure~\ref{fig:rch4_correlation} shows such correlations between \rchtwo{}, \rntwo{}, and \rchfour{} for \ca{} (left) and \pb{} (right).
Each correlation plot shows individual points each corresponding to the IMSRG(2) predictions for a single Hamiltonian: \dnnlogo{} (orange cross), \arthuisem{} (green plus), and the 34 nonimplausible Hamiltonians (blue circles).

We find strong correlations between \rchtwo{} and \rchfour{} for both \ca{} and \pb{}.
While variations in nuclear Hamiltonians may allow for a broad range in predicted charge radii, such variations clearly produce correlated changes in charge density moments, not independent ones.
Past work has also shown that model-space and many-body truncation uncertainties are similarly correlated~\cite{Heinz2025, Heinz2024inprep_MuToEResponses}, which we also find for the \dnnlogo{} and \arthuisem{} Hamiltonians.
Specifically, for the \arthuisem{} (\dnnlogo{}) Hamiltonian, we find that \rchtwo{} and \rchfour{} of \pb{} change by 0.7\% and 1.2\% (0.8\% and 1.7\%), respectively, when we vary the model-space truncation from 15 to 13 major shells.
Additionally they change by 0.5\% and 0.9\% (0.3\% and 0.5\%), respectively, when we vary \ethreemax{} from 28 to 24.
In both cases, these shifts are perfectly along the established correlation, meaning that model-space uncertainties are correlated in the same way as Hamiltonian uncertainties.
While we do not explicitly test the IMSRG(2) truncation here, past calculations involving the IMSRG(3)~\cite{Heinz2021PRC_IMSRG3, Heinz2024inprep_MuToEResponses} have shown that for quantities related to nuclear densities the many-body corrections from the IMSRG(3) also follow the same established correlation.

We also find a strong, but less tight correlation between \rntwo{} and \rchfour{} for both systems.
The larger scattering of the nonimplausble Hamiltonians suggests that there is some uncertainty in nuclear Hamiltonians that allows for slight variations in the relative behavior of the proton and neutron densities.
Nonetheless, such variations are limited, and there is a strong microscopic connection between \rchfour{} and \rntwo{} that can be exploited to study the neutron radius without directly measuring it, as suggested in Refs.~\cite{Kurasawa2019,Kurasawa2021PTEP_R4Measurement,Kurasawa2022PTEP}.
This is closely related to the established connection between \rchtwo{} and \rntwo{}.

The large Hamiltonian uncertainty reflected by the broad range of \rchtwo{} predictions can be reduced by calibrating the theory predictions against additional constraints~\cite{Hu2022NP_Pb208, Heinz2024inprep_MuToEResponses}.
For example, by combining experimental \rchtwo{} values~\cite{Noel2024JHEP_ChargeDensities,Kurasawa2021PTEP_R4Measurement} with our correlations, we provide a prediction for the value of \rchfour{}, indicated by the gray bands.
For each correlation, we include a conservative error band such that all the predictions from nonimplausible Hamiltonians lie within the band.
Combining this with the experimental uncertainties, we predict $\rchfour (\ca) = 204.4 \pm 3.2\:\fm^4$, $\rchfour (\pb) = 1172.5 \pm 16.7\:\fm^4$.
These values are consistent with the experimental values extracted in Ref.~\cite{Kurasawa2021PTEP_R4Measurement}. 
Note that the extracted value $\rchfour (\ca) = 194.734 \pm 2.544\:\fm^4$ disagrees with our prediction because the \rchtwo{} value used in that work is also smaller than what was found in a more recent analysis in Ref.~\cite{Noel2024JHEP_ChargeDensities} ($11.910 \pm 0.061\:\fm^2$ vs.\ $12.076 \pm 0.070\:\fm^2$).
In \pb{}, the value extracted in \cite{Kurasawa2021PTEP_R4Measurement}, $\rchfour (\pb) = 1171.981 \pm 17.627\:\fm^4$, is in excellent agreement with our prediction.

We also perform a similar analysis using the more uncertain correlation between \rntwo{} and \rchfour{} and theory predictions for \rntwo{}.
Based on a previous prediction for \rntwo{}~\cite{Heinz2024inprep_MuToEResponses}, we obtain $\rchfour (\ca) = 205.8 \pm 10.7\:\fm^4$,
which is in excellent agreement with the result inferred from the \rchtwo{}--\rchfour{} correlation, but with a larger uncertainty.

\section{Implications for neutron skins}

The correlation between \rchfour{} and \rntwo{} is the basis for extracting information about the neutron density from high-precision electron scattering.
The \rntwo{}--\rchfour{} correlation is however substantially more uncertain than the \rchtwo{}--\rchfour{} correlation (and the closely related \rptwo{}--\rchfour{} correlation).
We see the impact of this uncertainty in Fig.~\ref{fig:skin_correlation},
where we no longer observe a significant correlation between $\rskin = (\rntwo)^{1/2} - (\rptwo)^{1/2}$ and \rchfour{} in either \ca{} or \pb{}.
Without such a correlation, we cannot extract a value for \rskin{} from our \rchfour{} predictions.
A broad range of neutron skin predictions could be compatible with the constraints coming from \rchfour{},
consistent with a large set of independent determinations and predictions of \rskin{} in \ca{} and \pb{}~\cite{Adhikari2022, Birkhan2017PRL, Hagen2016NP_Ca48Skin, Hu2022NP_Pb208, Heinz2024inprep_MuToEResponses, Reinhard2022PRL,Adhikari2021, Tamii2011PRL, Klos2007, Zenihiro2010,
Reinhard2021PRL, Kurasawa2021PTEP_R4Measurement}.

This suggests that while high-precision electron scattering is sensitive to the neutron radius, determining \rskin{} from such a measurement will be much more challenging.
In particular, Ref.~\cite{Kurasawa2021PTEP_R4Measurement} found substantial model dependence in the extraction of \rntwo{} from \rchfour{} leading to substantial model dependence in the inferred \rskin{}.
Our predictions do not suffer from the same model dependence as phenomenological methods, and because of this we do not find spurious correlations between \rskin{} and \rchfour{}.
We conclude that inferring \rskin{} from a measurement of \rchfour{} will have some inherent model dependence.
It is still possible that as nuclear Hamiltonians and the charge density operator are further improved, for example, by reaching higher orders in the EFT expansion or through more sophisticated global optimizations, we find that this picture evolves.
From this perspective, precision measurements of electroweak properties of nuclei are important for guiding the development and optimization of both nuclear forces and currents.
Additionally, precise measurements of \rchtwo{} and \rchfour{} do constrain the neutron radii of nuclei when combined with the strong correlations between these observables we find.

\section{Conclusions}

We calculated the ground states of \ca{} and \pb{} with ab initio methods using a large ensemble of state-of-the-art nuclear Hamiltonians from chiral EFT.
We computed point-proton and point-neutron radii and the charge form factors $\fch(q^2)$, used to extract the second- and fourth-order moments of the charge density.
We found that this extraction was most reliable when using Gaussian processes to interpolate $\fch(q^2)$ for small $q$, which also allowed us to quantify and control the uncertainty associated with the interpolation.
Based on our computed values for \rchtwo{}, \rchfour{}, \rntwo{}, and \rskin{}, we established strong microscopic correlations between \rchfour{} and both \rchtwo{} and \rntwo{}.
We leveraged these correlations to provide predictions for \rchfour{} in \ca{} and \pb{}.
When considering \rchfour{} and \rskin{}, we found only a weak correlation, insufficient to extract \rskin{} from \rchfour{}.
This limits the ability of high-precision electron scattering to constrain the neutron skin of these systems in a model-independent way.
On the other hand, high-precision isotope-shift spectroscopy is sensitive to changes in \rchfour{} across isotope chains and can give insight into the evolution of deformation~\cite{Door2024arxiv_YbBoson}.
High-precision electron scattering can provide reference values for \rchfour{} for such studies, possibly also constraining nuclear structure effects for searches for new physics~\cite{Hur2022PRL_YbIS, Door2024arxiv_YbBoson, Wilzewski:2024wap}.

\section*{Data availability}

The data shown in the figures and table of this work are available in Ref.~\cite{miyagi_2025_17254564}.

\section*{Acknowledgments}

We thank Pierre Arthuis for fruitful discussions.
This work was supported by JST ERATO Grant No.~JPMJER2304, Japan, by JSPS KAKENHI Grant Numbers 25K07294, 25K00995, and 25K07330,
by the U.S.\ Department of Energy, Office of Science, Office of Advanced Scientific Computing Research and Office of Nuclear Physics, Scientific Discovery through Advanced Computing (SciDAC) program (SciDAC-5 NUCLEI), 
by the Laboratory Directed Research and Development Program of Oak Ridge National Laboratory, managed by UT-Battelle, LLC, for the U.S.\ Department of Energy,
and by the European Research Council (ERC) under the European Union's Horizon 2020 research and innovation programme (Grant Agreement No.~101020842).
This research used resources provided by Multidisciplinary Cooperative Research Program in Center for Computational Sciences, University of Tsukuba, by the Gauss Centre for Supercomputing e.V.~(www.gauss-centre.eu) through the John von Neumann Institute for Computing (NIC) on JUWELS at J\"{u}lich Supercomputing Centre (JSC), and of the Oak Ridge Leadership Computing Facility located at Oak Ridge National Laboratory, which is supported by the Office of Science of the Department of Energy under contract No.~DE-AC05-00OR22725.

This manuscript has been authored in part by UT-Battelle, LLC, under contract DE-AC05-00OR22725 with the US Department of Energy (DOE). The US government retains and the publisher, by accepting the article for publication, acknowledges that the US government retains a nonexclusive, paid-up, irrevocable, worldwide license to publish or reproduce the published form of this manuscript, or allow others to do so, for US government purposes. DOE will provide public access to these results of federally sponsored research in accordance with the DOE Public Access Plan (\url{http://energy.gov/downloads/doe-public-access-plan}).

\appendix

\section{Extracting moments from charge form factors}
\label{app:moments}

\begin{figure*}
    \centering
    \includegraphics[width=\textwidth]{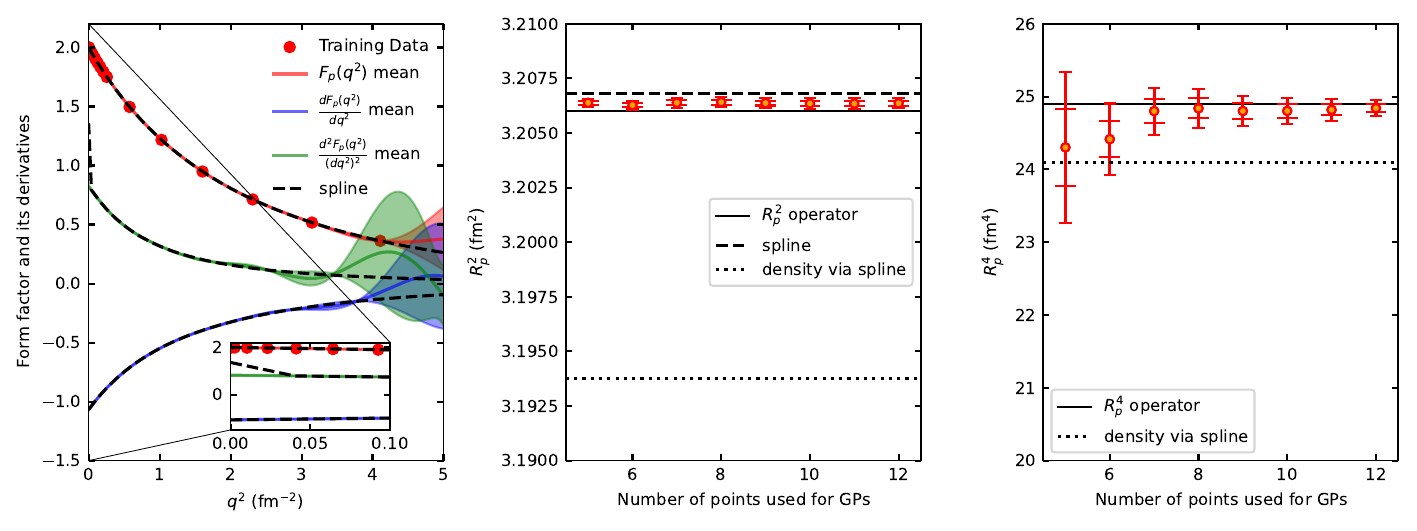}
\caption{Point-proton form factor (left), point-proton mean-squared radius \rptwo{} (middle), and fourth-order moment of point-proton density distribution \rpfour{} (right) for $^{3}$He. In the left panel, the form factor and its first and second derivatives are shown. The circles are points used to train the GP, whose 95\% confidence intervals are given by shaded areas. The dashed curves are computed with cubic spline interpolation. In the middle panel, \rptwo{} are computed with the GPs trained with the $n$ lowest $q^{2}$ points, and the $68\%$ and $95\%$ confidence intervals are given by the bars. The solid, dashed, and dotted lines are computed with the \rptwo{} operator, slope of the spline interpolation, and the radial integral of the density obtained with the spline interpolation, respectively. The right panel is the same except for \rpfour{}.}
\label{fig:GP2}
\end{figure*}

We validate our approach to computing \rchtwo{}, \rchfour{} from \fchint{} in $^3$He using the Jacobi no-core shell model (NCSM)~\cite{Barrett2013PPNP_NCSMReview} to compute the ground state.
The Jacobi NCSM calculations are performed with the 1.8/2.0~(EM) interaction~\cite{Hebeler2011PRC_SRG3NFits} at $N_{\rm max}=30$ and $\hbar\omega=20$ MeV.
To keep the comparison simple, we focus on the second- and fourth-order moments of the point-proton density \rptwo{} and \rpfour{}, respectively. 
As the Jacobi NCSM does not contain the center-of-mass coordinate by construction, \rptwo{} and \rpfour{} can be evaluated with the corresponding one-body operator.
This is not generally true because the intrinsic \rptwo{} and \rpfour{} operators include up to two- and four-body terms, respectively~\cite{Door2024arxiv_YbBoson}. 

For this test, we evaluate the intrinsic point-proton form factor $F_{\rm p}(q^{2})$ at $q = 0$, 1, 5, 10, 20, 30, 40, 50, 60, 70, 80, 90, 100, 150, 200, 250~MeV.
We interpolate $F_{\rm p}(q^{2})$ using a simple spline interpolation and using a Gaussian process (GP). 
Since derivatives of a GP are also GPs, it is straightforward to quantify the uncertainties of the derivatives. 
Following Ref.~\cite{Drischler2020PRC_NuclearMatterGPBLong}, we construct the GPs with \texttt{scikit-learn}~\cite{scikit-learn}.
We also compute the radii based on Eq.~\eqref{eq:Rn-dens} through $F_{\rm p}(q^{2})$, where the calculations are extended to $q= 900$ MeV.

In Fig.~\ref{fig:GP2}, we explore these results to establish consistency between the different methods used: calculations based on operator expectation values, via Fourier transforms of the density, and via differentiation at $q^2\to0$.
In the left panel, we show the computed values of $F_{\rm p}(q^{2})$ as red points, together with the interpolating spline (dashed line) and the interpolating GP (red line and band).

Differentiating $F_\mathrm{p}(q^{2})$ once gives \rptwo{}, summarized in the middle panel.
We find that all methods are relatively consistent within 0.4\%, with the derivatives of the spline interpolator and the GP interpolator agreeing very well with the \rptwo{} operator expectation value.
The GP interpolator, however, provides an uncertainty estimate, and this uncertainty is already small and not changed much by including the higher $q^2$ values to train the GP.
The larger difference for \rptwo{} obtained via the density as computed as a Fourier transform of the spline interpolator is still small, less than $ 0.4\%$.
This comes from evaluating $F_{\rm p}(q^{2})$ at too few $q^2$ values to obtain the density to higher accuracy.

Differentiating $F_{\rm p}(q^{2})$ twice gives \rpfour{}, summarized in the right panel.
Here we find that the GP interpolator gives significantly larger relative uncertainties when trained with a small number of points than for \rptwo{}.
For larger numbers of training points, we find that the values computed from the GP interpolator and the \rpfour{} operator agree excellently.
The value computed from the density also agrees with the operator expectation value within 4\%.
On the other hand, the second derivative of the spline interpolator gives a completely different result, 40.85 fm$^{4}$ and not shown in the figure, making it unusable for extracting \rpfour{}.
This suddenly large second derivative can also be seen in the left panel for $q^2\to0$ in the inset.

\balance
\bibliographystyle{apsrev4-1-mh-mod}
\bibliography{ref}

\end{document}